\documentclass[12pt,twoside,fleqn]{article}

\usepackage{a4wide,cite}
\usepackage{amsmath,amssymb} 
\usepackage{graphicx}
\usepackage{color}

\numberwithin{equation}{section}
\allowdisplaybreaks[4]

\newcommand{\be}{\begin{equation}}
\newcommand{\ee}{\end{equation}}
\newcommand{\bea}{\begin{eqnarray}}
\newcommand{\eea}{\end{eqnarray}}

\newcommand{\MS}{\ensuremath{\overline{\text{MS}}}}

\begin{document}

\begin{titlepage}

\flushright{DESY 14-034}

\vspace{-0.5cm}

\flushright{March 2014}

\vspace{1cm}

\centerline{\Large\bf NLL QCD contribution of the electromagnetic dipole}
\vspace{2mm}
\centerline{\boldmath \Large\bf  operator to
$\bar{B}\to X_s\gamma \gamma$ with a massive strange quark}
\vspace{2.5cm}

\begin{center}
  {\bf H.M.~Asatrian$^a$ and C.~Greub$^b$}\\[1mm]
  {$^a$\sl Yerevan Physics Institute, 0036 Yerevan, Armenia.}\\[1mm]
  {$^b$\sl Albert Einstein Center for Fundamental Physics, Institute for Theoretical Physics,\\
    Univ.~of Bern, CH-3012 Bern, Switzerland and}\\[1mm]
  {\sl Theory Group, Deutsches Elektronen-Synchrotron DESY, D-22603
    Hamburg, FRG.}\\[1mm]
 
\end{center}

\medskip

\vspace{2cm}
\begin{abstract}
\noindent 
We calculate the $O(\alpha_s)$ corrections to the double 
differential decay width 
$d\Gamma_{77}/(ds_1 \, ds_2)$
for the process $\bar{B} \to X_s \gamma \gamma$ originating from
diagrams involving the electromagnetic dipole operator
${\cal O}_7$. The kinematical variables $s_1$ and $s_2$ are defined as
$s_i=(p_b - q_i)^2/m_b^2$, where $p_b$, $q_1$, $q_2$ are the momenta
of $b$-quark and two photons. 
We introduce a nonzero mass $m_s$ for the strange quark to regulate
configurations where the gluon or one of the photons become collinear
with the strange quark and retain terms which are logarithmic in
$m_s$, while discarding terms which go to zero in the limit $m_s \to 0$.
When combining virtual- and bremsstrahlung corrections, the infrared
and collinear singularities induced by soft and/or collinear gluons
drop out. By our cuts the photons do not become soft, but one of them can
become collinear with the strange quark. This implies that in the
final result a single logarithms of $m_s$ survives. In principle
the configurations with collinear photon emission could be treated
using fragmentation functions. In a related work we found that similar
results can be obtained when simply interpreting $m_s$ appearing in
the final result as a constituent mass. We do so in the present paper
and vary $m_s$ between $400$ MeV and $600$ MeV in the numerics. 
This work extends a previous paper of us,
where only the leading power terms w.r.t. the (normalized)
hadronic mass $s_3=(p_b-q_1-q_2)^2/m_b^2$ were taken into account 
in the underlying triple differential decay width $d\Gamma_{77}/(ds_1 ds_2 ds_3)$.

\end{abstract}

\end{titlepage}

\section{Introduction}
Inclusive rare $B$-meson decays are known to be a unique source of indirect information about 
physics at scales of several hundred GeV. In the Standard Model (SM) all these processes 
proceed through loop diagrams and thus are relatively suppressed. In the extensions 
of the SM the contributions stemming from the diagrams with ``new'' 
particles in the loops can be comparable or even larger than the contribution from 
the SM. Thus getting experimental information on rare decays puts strong 
constraints on the extensions of the SM or can even lead to a  
disagreement with the SM predictions, providing evidence for some ``new physics''. 

To make a rigorous comparison between experiment and theory, precise
SM calculations for the (differential) decay rates are mandatory. While the
branching ratios for $\bar{B} \to X_s \gamma$ \cite{Misiak:2006zs}
and $\bar{B} \to X_s \ell^+
\ell^-$ are known today even to
next-to-next-to-leading logarithmic (NNLL) precision (for reviews, see
\cite{Hurth:2010tk,Buras:2011we}),
other branching ratios, like the one for $\bar{B} \to X_s \gamma
\gamma$ discussed in this paper, are systematically only known to leading logarithmic
(LL) precision in the SM
\cite{Simma:1990nr,Reina:1996up,Reina:1997my,Cao:2001uj}. In
\cite{Asatrian:2011ta} the NLL result for the contribution associated
with the photonic dipole operator ${\cal O}_7$ was worked out for
$\bar{B} \to X_s \gamma \gamma$ in a certain approximation (details see below). 
In contrast to $\bar{B} \to X_s
\gamma$, the current-current operator ${\cal O}_2$ has a non-vanishing matrix
element for $b \to s \gamma \gamma$ at order $\alpha_s^0$ precision, 
leading to an interesting interference pattern with the contributions associated
with the electromagnetic dipole operator ${\cal O}_7$ already at LL
precision. As a consequence, potential new physics should be clearly visible
not only in the total branching ratio, but also in the 
differential distributions.
 
As the process $\bar{B} \to X_s \gamma \gamma$ is expected to be measured at
the planned Super $B$-factories, it is necessary
to calculate the differential distributions to NLL precision in the
SM, in order to
fully exploit its potential concerning new physics. 
The starting point of our calculation is the effective Hamiltonian,
obtained by integrating out the heavy particles in the SM, leading to
\be
 {\cal H}_{eff} = - \frac{4 G_F}{\sqrt{2}} \,V_{ts}^\star V_{tb} 
   \sum_{i=1}^8 C_i(\mu) {\cal O}_i(\mu)  \, ,
\label{Heff}
\ee
where we use the operator basis introduced in \cite{Chetyrkin:1996vx}:
\be
\begin{array}{llll}
{\cal O}_1 \,= &\!
 (\bar{s}_L \gamma_\mu T^a c_L)\, 
 (\bar{c}_L \gamma^\mu T_a b_L)\,, 
               &  \quad 
{\cal O}_2 \,= &\!
 (\bar{s}_L \gamma_\mu c_L)\, 
 (\bar{c}_L \gamma^\mu b_L)\,,   \\[1.002ex]
{\cal O}_3 \,= &\!
 (\bar{s}_L \gamma_\mu b_L) 
 \sum_q
 (\bar{q} \gamma^\mu q)\,, 
               &  \quad 
{\cal O}_4 \,= &\!
 (\bar{s}_L \gamma_\mu T^a b_L) 
 \sum_q
 (\bar{q} \gamma^\mu T_a q)\,,  \\[1.002ex]
{\cal O}_5 \,= &\!
 (\bar{s}_L \gamma_\mu \gamma_\nu \gamma_\rho b_L) 
 \sum_q
 (\bar{q} \gamma^\mu \gamma^\nu \gamma^\rho q)\,, 
               &  \quad 
{\cal O}_6 \,= &\!
 (\bar{s}_L \gamma_\mu \gamma_\nu \gamma_\rho T^a b_L) 
 \sum_q
 (\bar{q} \gamma^\mu \gamma^\nu \gamma^\rho T_a q)\,,  \\[1.002ex]
{\cal O}_7 \,= &\!
  \frac{e}{16\pi^2} \,\bar{m}_b(\mu) \,
 (\bar{s}_L \sigma^{\mu\nu} b_R) \, F_{\mu\nu}\,, 
               &  \quad 
{\cal O}_8 \,= &\!
  \frac{g_s}{16\pi^2} \,\bar{m}_b(\mu) \,
 (\bar{s}_L \sigma^{\mu\nu} T^a b_R)
     \, G^a_{\mu\nu}\, .
\end{array} 
\label{opbasis}
\ee
The symbols $T^a$ ($a=1,8$) denote the $SU(3)$ color generators; 
$g_s$ and $e$, the strong and electromagnetic coupling constants.
In eq.~(\ref{opbasis}), 
$\bar{m}_b(\mu)$ is the running $b$-quark mass 
in the $\MS$-scheme at the renormalization scale $\mu$.
As we are not interested in CP-violation effects in the present paper, we 
made use of the approximation
 $V_{ub} V_{us}^* \ll V_{tb} V_{ts}^* $ when writing
eq. (\ref{Heff}). We also put the mass of the strange quark to zero
which in principle enters ${\cal O}_7$, because in this paper we will
work out only terms which are logarithmic in $m_s$ or independent of $m_s$. 
  
While the Wilson coefficients $C_i(\mu)$ appearing in eq. (\ref{Heff})
are known to sufficient precision at the low scale $\mu \sim m_b$
since a long time (see e.g. the reviews \cite{Hurth:2010tk,Buras:2011we}
and references therein), the matrix elements 
$\langle s \gamma \gamma|{\cal  O}_i|b\rangle$ and 
$\langle s \gamma \gamma \, g|{\cal  O}_i|b\rangle$, 
which in a NLL calculation are needed to order
$g_s^2$ and $g_s$, respectively, are not known yet. To calculate the
$({\cal O}_i,{\cal O}_j)$-interference contributions for the
differential distributions at order
$\alpha_s$ is in many respects of similar complexity as the
calculation of the photon energy spectrum in $\bar{B} \to X_s \gamma$ 
at order $\alpha_s^2$
needed for the NNLL computation. There, the individual
interference contributions, which all involve extensive calculations, were
published in separate papers, sometimes even by two independent groups
(see e.g. \cite{Melnikov:2005bx} and \cite{Asatrian:2006sm}).
It therefore cannot be expected that the NLL results for the
differential distributions related to $\bar{B} \to X_s \gamma \gamma$ are 
given in a single paper. 

As a first step towards a NLL prediction for $\bar{B} \to X_s \gamma
\gamma$, we calculated in 2011 the $O(\alpha_s)$ corrections 
to the $({\cal O}_7,{\cal O}_7)$-interference contribution to the double 
differential decay width $d\Gamma/(ds_1 ds_2)$ at the partonic level,
using an approximation where only the
leading power w.r.t. the (normalized) hadronic mass 
were retained in the underlying triple differential decay width
$d\Gamma_{77}/(ds_1 ds_2 ds_3)$ \cite{Asatrian:2011ta}. 
The variables $s_1$ and $s_2$ are defined as $s_i=(p_b-q_i)^2/m_b^2$, where $p_b$
and $q_i$ denote the four-momenta of the $b$-quark and the two
photons, respectively and $s_3$ denotes the normalized hadronic mass
of the final state, i.e. $s_3=(p_b-q_1-q_2)^2/m_b^2$.

At order $\alpha_s$
there are contributions to $d\Gamma_{77}/(ds_1 ds_2)$ with three
particles ($s$-quark and two photons) in the final state and a gluon
in the loop [virtual corrections] and tree-level contributions with
four particles ($s$-quark, two photons and a gluon) in the final
state [bremsstrahlung corrections].

As we will discuss in section \ref{sec:leadingorder}, we work out the QCD corrections
to the double differential decay width 
in the kinematical range 
\[
0 < s_1 < 1 \quad ; \quad 0 < s_2 < 1-s_1 \, .
\]
Concerning the virtual corrections, all singularities (after
ultra-violet renormalization) are due to  soft gluon exchange
and/or  collinear gluon exchange involving the $s$-quark. Concerning the
bremsstrahlung corrections (restricted to the same range of $s_1$ and
$s_2$), there are also singularities due to soft- and/or collinear
gluons, but there are additional kinematical situations where one of the
photons is emitted collinear to the $s$-quark. While the
singularities induced by gluons cancel when combining virtual- and
bremsstrahlung corrections, those associated with collinear photons
remain, as discussed in detail in section \ref{sec:bremsstrahlung}. 
In ref. \cite{Asatrian:2011ta} we found, however, that there are no
singularities associated with collinear photon emission in the double
differential decay width when only retaining
the leading power w.r.t. to the (normalized) hadronic mass
$s_3=(p_b - q_1 - q_2)^2/m_b^2$ in the underlying triple differential 
distribution $d\Gamma_{77}/(ds_1 ds_2 ds_3)$. The results in
ref. \cite{Asatrian:2011ta} were obtained within this
``approximation''.  

The main goal of the present paper is to go beyond this approximation.
When doing so, the singularities induced by collinear photon emission
from the strange quark remain in the final perturbative result and
additional concepts like parton
fragmentation functions of a quark into a photon are
needed \cite{Kapustin:1995fk}. In our recent work \cite{Asatrian:2013raa}
on the tree-level
contributions of the operators $O^u_{1,2}$ to the branching ratio for
the process $\bar{B} \to X_d \gamma$, we found
that the results involving fragmentation functions are similar to
those obtained by providing the quark $q$ which radiates an (almost) collinear
photon with an appropriately chosen constituent mass $m_q$. The
approach using constituent masses was also used in
ref. \cite{Kaminski:2012eb}, where the analogous contributions to
$\bar{B} \to X_s \gamma$ were investigated. 

As the approach with a constituent mass is technically easier and,
more importantly, because the fragmentation functions are not known 
accurately as discussed in \cite{Asatrian:2013raa}, we interpret 
$m_s$, which we originally introduce as a regulator of collinear
singularities, as a constituent mass in the present paper and retain all terms
of the type $\log^n(m_s)$, while neglecting power terms in $m_s$, as
well as terms of the form $m_s^n \log^m(m_s)$, which tend to zero in the
limit $m_s \to 0$. As the virtual- and bremsstrahlung corrections in
\cite{Asatrian:2011ta} were calculated for a massless strange quark
(which means dimensional regularization of collinear
singularities), we have to redo both parts in the present work. 
  
Before moving to the detailed organization of our paper, we should
mention that the inclusive double radiative process $\bar{B} \to X_s \gamma
\gamma$ has also been explored in several extensions of the SM
\cite{Reina:1996up,Gemintern:2004bw,Cao:2001uj}. Also 
the corresponding exclusive modes, $B_s \to \gamma \gamma$  
and $B\to K \gamma \gamma$, have been 
examined before, both in the SM
\cite{Reina:1997my,Chang:1997fs,Hiller:1997ie,Bosch:2002bv,Bosch:2002bw,
Hiller:2004wc,Hiller:2005ga,Lin:1989vj,Herrlich:1991bq,Choudhury:2002yu}
and in its extensions 
 \cite{Aliev:1997uz,Hiller:2004wc,Hiller:2005ga,Bertolini:1998hp,Gemintern:2004bw,Bigi:2006vc,
Devidze:1998hy,Aliev:1993ea,Xiao:2003jn,XiuMei:2011iv,Huo:2003cj,Chen:2011te}.
We should add that the long-distance resonant effects were
also discussed in the literature (see e.g. \cite{Reina:1997my} and the
references therein). 
Finally, the effects of photon emission from the spectator quark in
the $B$-meson were discussed in \cite{Chang:1997fs,Hiller:2004wc,Ignatiev:2003qm}. 

The remainder of this paper is organized as follows.
In section \ref{sec:leadingorder} we work out the double differential distribution 
$d\Gamma_{77}/(ds_1 ds_2)$ in leading order, i.e., without taking into
account QCD corrections to the matrix element 
$\langle s \gamma \gamma |{\cal  O}_7|b\rangle$. 
In this section we also give the order $\alpha_s^0$ results when
including the effects of the operators ${\cal O}_1$ and ${\cal O}_2$.
Section \ref{sec:virtual} is devoted to the calculation of the virtual  
corrections of order $\alpha_s$ to the double differential decay
width in a scheme where the collinear singularities are regulated
using a nonzero strange quark mass $m_s$. 
In section \ref{sec:bremsstrahlung}
the corresponding gluon bremsstrahlung corrections to the double
differential width are worked out. 
In section \ref{sec:combination} virtual- and bremsstrahlung corrections are combined and
the result for the double differential decay width is given. As our
analytic results (in particular those for the bremsstrahlung
corrections) are rather lengthy, we
prefer to give certain parts of our results in the form of fits
which involve simple ``basis functions''. 
In section \ref{sec:numerics} we illustrate the numerical impact of the
NLL corrections. 
A comparison with the results in \cite{Asatrian:2011ta}, where 
only the leading power w.r.t. the (normalized)
hadronic mass $s_3$ was retained at the level of the triple
differential decay width $d\Gamma_{77}/(ds_1 ds_2 ds_3)$, is also done
in this section. The main text of our paper ends with a short summary in section
\ref{sec:summary}. The appendices \ref{appendixvirt},
\ref{sec:phasespacemassive} and \ref{append:renormalizationconstants}
contain intermediate results and technical ingredients.

\section{Leading order result}\label{sec:leadingorder}
\begin{figure}[h]
\begin{center}
\includegraphics[width=0.9\textwidth]{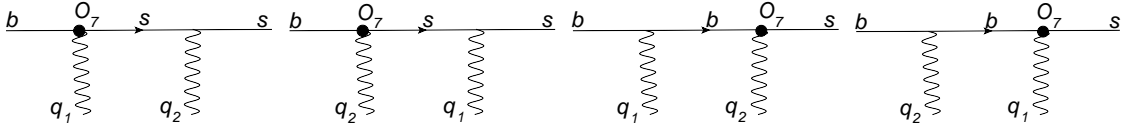}
\caption{The diagrams defining the tree-level
amplitude for $b \to s \gamma \gamma$ associated with ${\cal O}_7$ are
shown. The four-momenta of the $b$-quark, the $s$-quarks and the two
photons are denoted by $p_b$, $p_s$, $q_1$ and $q_2$, respectively.}
\label{fig:amplitudetree}
\end{center}
\end{figure}

\begin{figure}[h]
\begin{center}
\includegraphics[width=5.0cm]{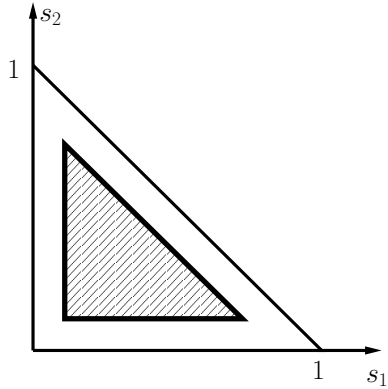}
\caption{The relevant phase-space region for $(s_1,s_2)$ used in this paper is 
shown by the shaded area.}
\label{fig:phasespace}
\end{center}
\end{figure}

In this section we discuss the double differential decay width
$d\Gamma_{77}/(ds_1 ds_2)$ at lowest order in QCD, i.e. $\alpha_s^0$.
The dimensionless variables $s_1$ and $s_2$ are defined everywhere in this
paper as
\be
s_1=\frac{(p_b-q_1)^2}{m_b^2} \quad ; \quad
s_2=\frac{(p_b-q_2)^2}{m_b^2} \, .
\label{kinematicvariables}
\ee
At lowest order the double differential decay width is based on the 
diagrams shown in Fig.~\ref{fig:amplitudetree}.
The variables
$s_1$ and $s_2$ form a complete set of kinematically independent
variables for the three-body decay $b \to s \gamma \gamma$.
Their kinematical range is as follows:
\[
0 \le s_1 \le 1 \quad ; \quad 0 \le s_2 \le 1-s_1 \, .
\]
The energies $E_1$ and $E_2$ in the rest-frame of the $b$-quark of the two photons are 
related to $s_1$ and $s_2$ in a simple way: $s_i=1-2 \, E_i/m_b$.
As the energies $E_i$ of the photons have to be away from zero in order to
be observed, the values of $s_1$ and $s_2$ can be considered to be
smaller than one. By additionally requiring $s_1$ and $s_2$ to be larger than zero,
we exclude collinear photon emission from the $s$-quark, because
$(p_s+q_1)^2=(p_b-q_2)^2=s_2 \, m_b^2>0$ and 
$(p_s+q_2)^2=(p_b-q_1)^2=s_1 \, m_b^2>0$. Using these cuts, $m_s$ can
be safely put to zero at leading order.
It is also easy to
implement a lower cut on the invariant mass squared $s$ of the two 
photons by observing that $s=(q_1+q_2)^2=1-s_1-s_2$. To parametrize
all the mentioned conditions in terms of one parameter $c$ (with $c>0$),
one can proceed as suggested in \cite{Reina:1996up}:
\be
s_1 \ge c \, , \quad s_2 \ge c \, , \quad 1-s_1-s_2 \ge c \, .
\label{kinematical_cuts}
\ee   
Applying such cuts, the relevant phase-space region in the
$(s_1,s_2)$-plane is shown by the shaded area in
Fig.~\ref{fig:phasespace}. Our aim in this paper is to work out the
double differential decay width in this restricted area of the $s_1$ and
the $s_2$ variable also when discussing the gluon bremsstrahlung
corrections\footnote{In this case, the normalized invariant mass
squared $s$ of the two photons reads $s=1-s_1-s_2+s_3$, where
$s_3$ is the normalized hadronic mass squared. The condition
$1-s_1-s_2 \ge c$ then still eliminates two-photon configurations with
small invariant mass.}.
To exhibit the
singularity structure of the virtual corrections discussed in the
next section in a transparent way, it is useful to give the
leading-order spectrum
in $d=4-2\epsilon$ dimensions. We obtain
\begin{eqnarray}
&&\frac{d\Gamma_{77}^{(0,d)}}{ds_1 \, ds_2} = \frac{\alpha^2 \,
    \bar{m}_b^2(\mu) \, m_b^3 
\, |C_{7,eff}(\mu)|^2 \, G_F^2 \,
  |V_{tb} V_{ts}^*|^2 \,  Q_d^2}{1024 \, \pi^5} \, \left(
  \frac{\mu}{m_b}\right)^{4\epsilon}   r 
\label{treea}
\end{eqnarray}
with
\begin{eqnarray}
&& r=\frac{\left[r_0+\epsilon (r_1+r_2+r_3+r_4) \right]
   \left(1-s_1-s_2\right)}{\left(1-s_1\right)^2
   s_1 \left(1-s_2\right)^2 s_2} \, .
\label{treeb}  
\end{eqnarray} 
In $r$ we retained  terms of order $\epsilon^1$, while discarding terms of higher order.
The individual pieces $r_0,\ldots,r_4$ read  
\begin{eqnarray}
\nonumber r_0&=&-48 s_2^3 s_1^3+96 s_2^2 s_1^3-56 s_2 s_1^3+8
   s_1^3+96 s_2^3 s_1^2-192 s_2^2 s_1^2+112
   s_2 s_1^2-56 s_2^3 s_1+
   \\&& 112 s_2^2 s_1-96 s_2
   s_1+8 s_1+8 s_2^3+8 s_2
\label{funr0}
    \end{eqnarray}
     \begin{eqnarray}
\nonumber r_1&=&-16 s_2^2 s_1^3+16 s_2 s_1^3-16 s_2^3 s_1^2+48
   s_2^2 s_1^2-32 s_2 s_1^2+16 s_1^2+16 s_2^3
   s_1- \\&& 32 s_2^2 s_1-16 s_2 s_1+16 s_2^2
    \end{eqnarray}
   \begin{eqnarray}
 && r_2= - r_0 \, \log \left(s_1\right) \quad ; \quad
r_3= - r_0 \, \log \left(s_2\right) \quad ; \quad
r_4= - r_0 \, \log \left(1-s_1-s_2\right) \, .
    \end{eqnarray}
In eq. (\ref{treea}) the symbols  $\bar{m}_b(\mu)$ and  $m_b$
denote the mass of the $b$-quark in the $\overline{\rm{MS}}$-scheme
and in the on-shell scheme, respectively and
$C_{7,eff}(\mu)$ is the effective Wilson coefficient of the operator
${\cal O}_7$ at the low scale
$(\mu \sim m_b)$, which has an expansion in $\alpha_s$ as follows:
\be
C_{7,eff}(\mu) = C_{7,eff}^{0}(\mu) + \frac{\alpha_s(\mu)}{4\pi} \,
C_{7,eff}^{1}(\mu) \, .
\label{wilsonexpand}
\ee
This Wilson coefficient is known for a long time (see ref. \cite{Chetyrkin:1996vx} and
references therein). Note that in this section
only the lowest order part $C_{7,eff}^{0}$ of
$C_{7,eff}$ is needed in eq. (\ref{treea}), while in the following sections the
$C_{7,eff}^{1}$ piece has to be retained. 
 
In $d=4$ dimensions, the leading-order spectrum (in our restricted
phase-space) is obtained by simply
putting $\epsilon$ to zero, obtaining
\begin{eqnarray}
&&\frac{d\Gamma_{77}^{(0)}}{ds_1 \, ds_2} = \frac{\alpha^2 \, \bar{m}_b^2(\mu) \, m_b^3 \, |C_{7,eff}(\mu)|^2 \, G_F^2 \,
  |V_{tb} V_{ts}^*|^2 \,  Q_d^2}{1024 \, \pi^5} \, 
  \frac{(1-s_1-s_2)}{(1-s_1)^2 s_1 (1-s_2)^2 s_2} \, r_0 \, .
\label{treezero}
\end{eqnarray}

For completeness, we also list the order $\alpha_s^0$ result which
takes into account the remaining contributions of the operators ${\cal O}_1$,
${\cal O}_2$ and ${\cal O}_7$. Using $\hat{m}_c=m_c/m_b$, 
one gets \cite{Asatrian:2011zz,Cao:2001uj} when adapted to the
operator basis in eq. (\ref{opbasis})
\bea
&&\frac{d\Gamma^{(0)}_{\rm{remaining}}}{ds_1 ds_2} =
 \frac{\alpha^2 \, \, m_b^5 \, \, G_F^2 \,
  |V_{tb} V_{ts}^*|^2 }{1024 \, \pi^5} \times
\nonumber \\
&& \left\{
4 \, Q_u^4 \left( C_2(\mu) + \frac{4}{3} C_1(\mu) \right)^2
\frac{(s_1+s_2)}{(1-s_1-s_2)^2} \left| 1-s_1-s_2-4 \, \hat{m}_c^2 \,
\arcsin^2(z) \right|^2 \right. \nonumber \\
&&
\left. +  
16 \, Q_d \, Q_u^2 \left( C_2(\mu) + \frac{4}{3} C_1(\mu) \right) \,
C_{7,eff}(\mu) \,
\left( 1-s_1-s_2-4 \, \hat{m}_c^2 \,
\mbox{Re}\left( \arcsin^2(z) \right) \right) \right\} \, ,
\label{treeallzero}
\eea
where we identified $\bar{m}_b(\mu)$ with $m_b$ (which is correct at
lowest order). The argument of the $\arcsin$ function reads
$z=\sqrt{(1-s_1-s_2)/(4 \, \hat{m}_c^2)}$, where $\hat{m}_c^2$ is
tacitly understood to have a small negative imaginary part.

\begin{figure}[h]
\begin{center}
\includegraphics[width=\textwidth]{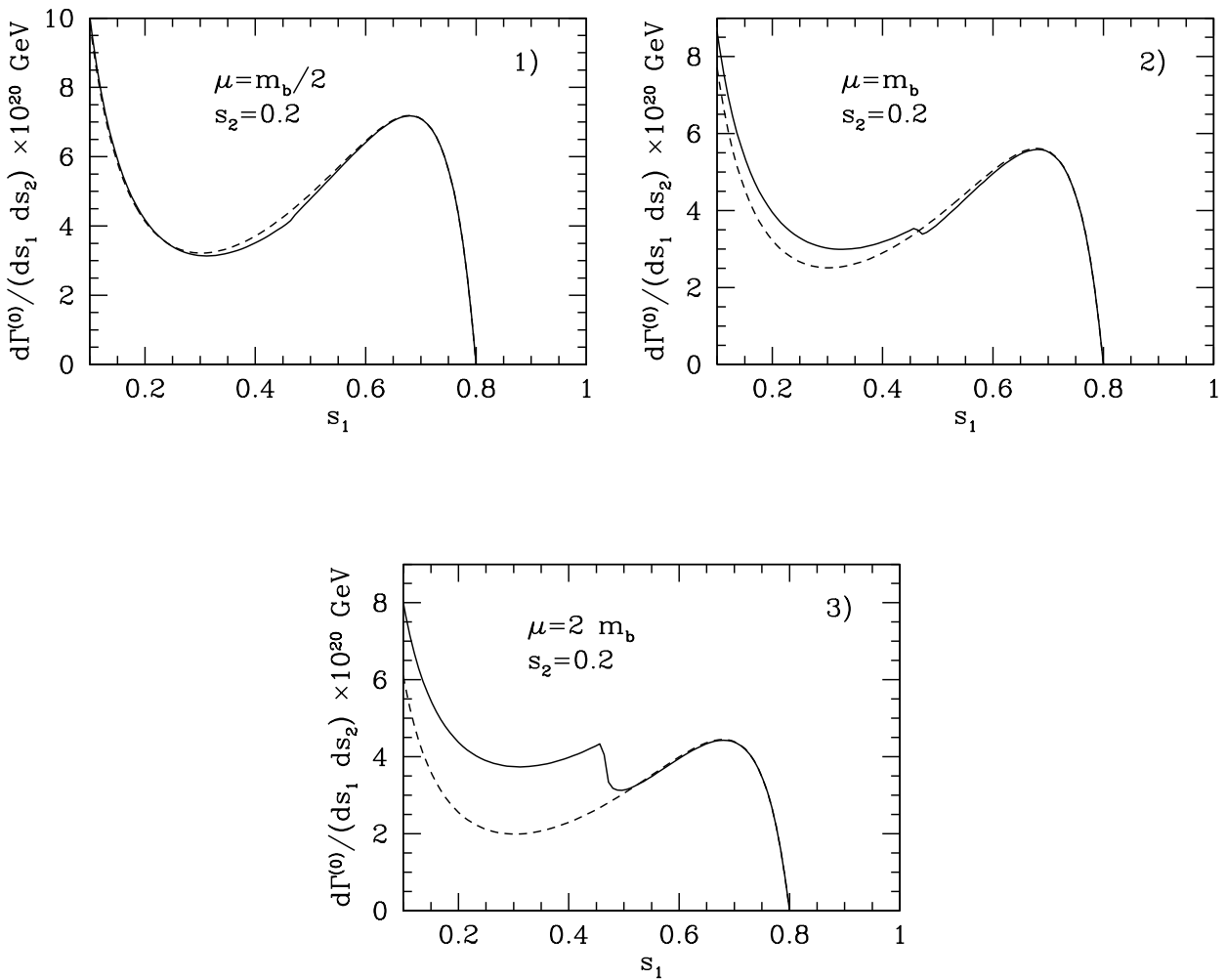}
\caption{Double differential decay width $d\Gamma^{(0)}/(ds_1 ds_2)$ at
leading order ($\alpha_s^0$) 
as a function of $s_1$ for $s_2$ fixed at $s_2=0.2$. 
The dashed line shows the result when only the $({\cal O}_7,{\cal
  O}_7)$ interference is taken into account, while the solid line
shows all contributions associated with ${\cal O}_1$, ${\cal O}_2$
and ${\cal O}_7$.
In the frames 1), 2) and 3) the renormalization scale is chosen to
  be $\mu=m_b/2$, $\mu=m_b$ and $\mu=2 \, m_b$, respectively.}
\label{fig:resultsLL}
\end{center}
\end{figure}

In Fig.~\ref{fig:resultsLL} we show the LL results based on eq. 
(\ref{treezero}) (dashed line) and the corresponding ones when also
including the contributions in eq. (\ref{treeallzero}) (solid line).
The numerical values of the input parameters and of the Wilson coefficients are
listed in tables \ref{tab:input} and \ref{tab:wilson}, respectively.
We see that for $\mu=m_b/2$ the $({\cal O}_7,{\cal O}_7)$ contribution
is by far the dominant
one. This can be easily understood from eq. (\ref{treeallzero}),
because the combination $\left( C_2(\mu)+\frac{4}{3} C_1(\mu) \right)$ is
almost zero at this scale. This is no longer true at $\mu=m_b$ or
$\mu=2m_b$, therefore the effects of the remaining terms become
more important.  

%
%
\begin{table}[h]
\centering \vspace{0.8cm}
\begin{tabular}{|c|c|}

 \hline
 Parameter & Value \\   \hline \hline

$m_{b}$& $4.8$~GeV   \\ \hline

$m_{c}/m_{b}$& $0.29$   \\ \hline

$m_{t}$&$175$~GeV    \\ \hline

$m_{W}$&$80.4$~GeV    \\ \hline

$m_{Z}$&$91.19$~GeV   \\ \hline

$G_{F}$&$1.16637\times10^{-5}$~\text{GeV}$^{-2}$  \\ \hline

$V_{tb} V_{ts}^* $&$0.04$    \\  \hline

$V_{cb}  $&$0.04$    \\  \hline

$\mbox{BR}_{sl}  $&$0.1049$    \\  \hline

${\alpha}^{-1}$&$137$    \\ \hline

${\alpha_{s}(M_{Z})}$&$0.119$      \\ \hline \hline
\end{tabular}
\caption{Values of the relevant input parameters} 
\label{tab:input}
\end{table}
\begin{table}[h]
\centering \vspace{0.8cm}
\begin{tabular}{|c|c|c|c|c|c|}
\hline
  &  $\alpha_s(\mu)$ & $C_{7,eff}^{0}(\mu)$ & $C_{7,eff}^{1}(\mu)$ &
$C_{1}^{0}(\mu)$ &  $C_{2}^{0}(\mu)$
\\   \hline \hline

$\mu= m_W$  & $0.1213$ & $-0.1957$ & $-2.3835$ & $0$ & $1$ \\ \hline 

$\mu=2 \, m_{b}$  & $0.1818$ & $-0.2796$ & $-0.1788$ & $-0.3352$ & $1.0116$ \\ \hline 

$\mu=m_{b}$  & $0.2175$ & $-0.3142$ &  $0.4728$ & $-0.4976$ & $1.0245$  \\ \hline

$\mu=m_{b}/2$  & $0.2714$ & $-0.3556$ & $1.0794$ & $-0.7117$ & $1.0478$   \\ \hline \hline

\end{tabular}
  \caption{$\alpha_s(\mu)$ and the Wilson coefficients
    $C_{7,eff}^{0}(\mu)$, $C_{7,eff}^{1}(\mu)$, $C_1^{0}(\mu)$, $C_2^{0}(\mu)$  at
    different values of the renormalization scale $\mu$.} 
\label{tab:wilson}
\end{table}
\section{Virtual corrections}\label{sec:virtual}
\begin{figure}[h]
\begin{center}
\includegraphics[width=0.9\textwidth]{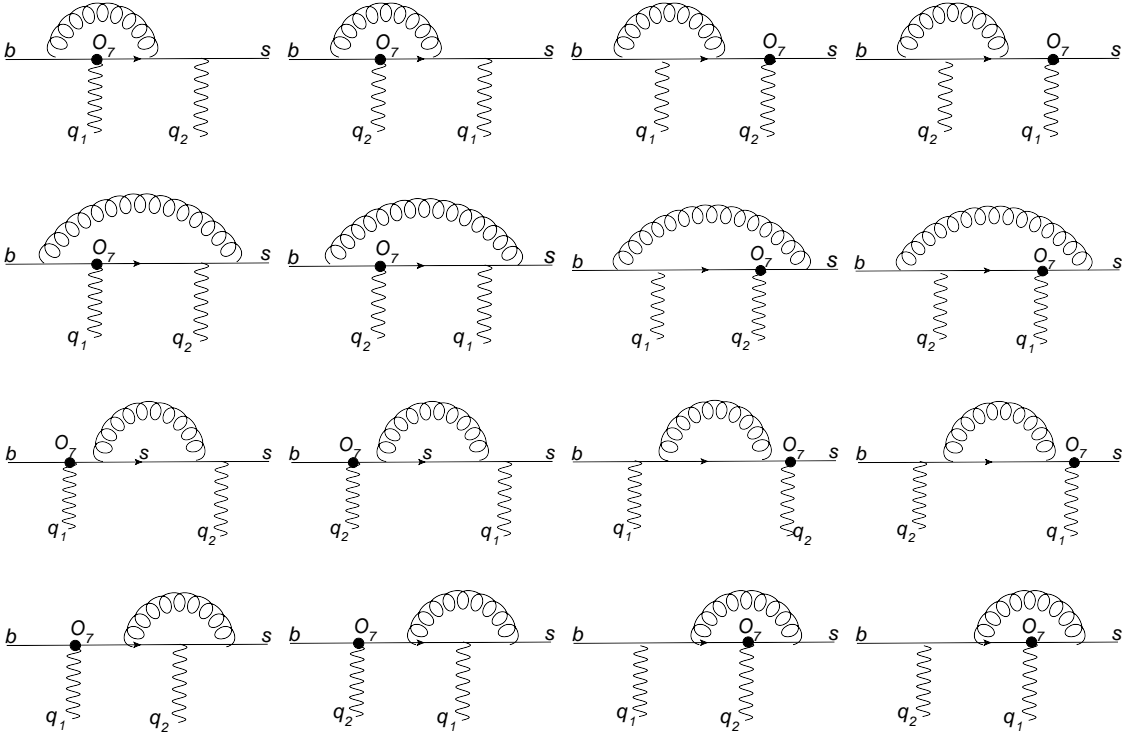}
\caption{The diagrams defining the one-loop
amplitude for $b \to s \gamma \gamma$ associated with ${\cal O}_7$ are
shown. Diagrams with self-energy insertions on the external quark-legs
are not shown.}
\label{fig:amplitudevirtual}
\end{center}
\end{figure}

We now turn to the calculation of the virtual QCD corrections, i.e. to
the contributions of order $\alpha_s$ with three particles in the
final state. The diagrams defining the (unrenormalized) 
virtual corrections at the amplitude level are shown in
Fig. \ref{fig:amplitudevirtual}. 
As the diagrams with a self-energy insertion on the external $b$- and
$s$-quark legs are taken into account in the renormalization process,
these diagrams are not shown in Fig. \ref{fig:amplitudevirtual}.
In order to get the (unrenormalized) virtual corrections 
$d\Gamma_{77}^{\rm bare}/(ds_1 ds_2)$ of order $\alpha_s$ to the
decay width, we have to work out the interference of the diagrams 
in Fig. \ref{fig:amplitudevirtual} with the leading
order diagrams in Fig. \ref{fig:amplitudetree}.

From the technical point of view, the calculation was made possible
by the use of the Laporta Algorithm \cite{Laporta:2001dd}  
(see also \cite{Tkachov:1981wb,Chetyrkin:1981qh})
to identify the
needed Master Integrals and by applying the differential equation method to
solve them. As we used these
techniques also in \cite{Asatrian:2011ta}, we refer
to section 7 of that paper which contains the technical details
and the corresponding references.
In appendix \ref{sec:phasespacemassive}
 we present, however, a technical issue which is specific
for the present work, viz. a useful parametrization of
the three-particle phase-space where one particle is massive.

In addition, we have to work out the counterterm contributions 
to the decay width. They can be split into two parts, according to
\be
\frac{d\Gamma_{77}^{\rm ct}}{ds_1 ds_2}=
\frac{d\Gamma_{77}^{{\rm ct},(A)}}{ds_1 ds_2}+
\frac{d\Gamma_{77}^{{\rm ct},(B)}}{ds_1 ds_2} \, .
\ee
\noindent
Part (A) involves the Lehmann, Symanzik, Zimmermann (LSZ) factors 
$\sqrt{Z_{2b}^{\rm OS}}$ and $\sqrt{Z_{2s}^{\rm OS}}$ for the $b$- and
$s$-quark field, as well as the
self-renormalization constant $Z_{77}^{\MS}$ of the operator ${\cal
  O}_7$ and $Z_{m_b}^{\MS}$ 
renormalizing the factor $\bar{m}_b(\mu)$ present in the
operator ${\cal O}_7$. The explicit results for these $Z$-factors are
given to relevant precision in Appendix \ref{append:renormalizationconstants}.
For part (A) we get
\be
\frac{d\Gamma_{77}^{{\rm ct},(A)}}{ds_1 ds_2} = 
\left[ \delta Z_{2b}^{\rm OS} + \delta Z_{2s}^{\rm OS} + 2 \, \delta Z_{m_b}^{\MS} +
  2 \, \delta Z_{77}^{\MS} \right] \, \frac{d\Gamma_{77}^{(0,d)}}{ds_1 ds_2} \, ,
\ee
where $d\Gamma_{77}^{(0,d)}/(ds_1 ds_2)$ is the leading order double
differential decay width in $d$-dimensions, as given in
eq. (\ref{treea}).

\begin{figure}[h]
\begin{center}
\includegraphics[width=0.6\textwidth]{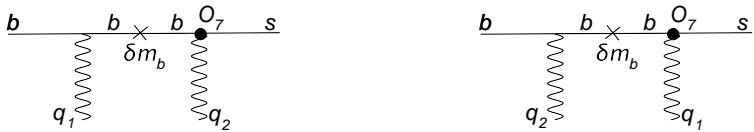}
\caption{Counterterm diagrams with a $\delta m_b$ insertion, see text.}
\label{fig:amplitudembinsertion}
\end{center}
\end{figure}
The counterterms defining part (B) are due to the insertion 
of $-i \delta m_b \bar b b$  
in the internal $b$-quark line in the leading order diagrams as
indicated in Fig. \ref{fig:amplitudembinsertion}, where 
\[
\delta m_b = (Z_{m_b}^{\rm OS}-1) \, m_b \, .
\]
More precisely, Part (B) consists of the interference of the diagrams in
Fig.~\ref{fig:amplitudembinsertion} with the leading order diagrams in 
Fig.~\ref{fig:amplitudetree}. When the strange quark is massive,
there is in principle also an analogous insertion of $-i \delta m_s 
\bar{s}s$ in internal $s$-quark lines. $\delta m_s$ is, however, proportional
to $m_s$ and since we neglect terms in which $m_s$ appears power-like,
we skip this contribution.

By adding $d\Gamma_{77}^{\rm bare}/(ds_1 ds_2)$ and 
$d \Gamma_{77}^{{\rm ct}}/(ds_1 ds_2)$,
we get the result for the renormalized virtual corrections to the
spectrum, $d\Gamma_{77}^{(1),virt}/(ds_1 \, ds_2)$.
It is useful to decompose this result into two pieces,
\begin{equation}
\frac{d\Gamma_{77}^{(1),virt}}{ds_1 \, ds_2} = 
\frac{d\Gamma_{77}^{(1,a),virt}}{ds_1 \, ds_2} + 
\frac{d\Gamma_{77}^{(1,b),virt}}{ds_1 \, ds_2}  \, .
\end{equation}
The infrared- and collinear singularities are completely contained in
$d\Gamma_{77}^{(1,a),virt}/(ds_1 \, ds_2)$. Explicitly, we obtain
(using $x_4=m_s^2/m_b^2$)
\begin{equation}
\frac{d\Gamma_{77}^{(1,a),virt}}{ds_1 \, ds_2} = \frac{\alpha_s}{4\pi}
\, C_F \,
\left[ \frac{4 \log (s_1+s_2)-4-2 \log(x_4)}{\epsilon} +
\log^2(x_4) -\log(x_4)
\right] \, \left( \frac{\mu}{m_b}\right)^{2\epsilon} \, \frac{d
  \Gamma_{77}^{(0,d)}}{ds_1 \, ds_2} 
\label{virtuala} 
\end{equation}
where  $d\Gamma_{77}^{(0,d)}/(ds_1 \, ds_2)$ is understood to be taken
exactly as given in eqs. (\ref{treea}) and (\ref{treeb}), i.e., by including the terms of
order $\epsilon^1$ in $r$. From the explicit expression  
$d\Gamma_{77}^{(1,a),virt}/(ds_1 \, ds_2)$
we see that the singularity structure consists of a simple
singular factor multiplying the corresponding tree-level decay width
in $d$-dimensions. We stress that the singularities (represented by
$1/\epsilon$ poles, $\log(x_4)$ terms and combinations thereof) 
are entirely due to
soft and/or collinear gluon exchange.
The infrared and collinear finite piece 
$d\Gamma_{77}^{(1,b),virt}/(ds_1 \, ds_2)$ can be written as
\begin{eqnarray}
&& \frac{d\Gamma_{77}^{(1,b),virt}}{ds_1 \, ds_2} = \frac{\alpha^2 \,  
\bar{m}_b^2(\mu) \, m_b^3 \, |C_{7,eff}(\mu)|^2 \, G_F^2 \,
  |V_{tb} V_{ts}^*|^2 \,  Q_d^2}{1024 \, \pi^5} \, \frac{\alpha_s}{4\pi} \, C_F 
\times  \nonumber \\
&&\hspace{2cm}  \left( 
\frac{-4 \, r_0 \, (1-s_1-s_2)}{(1-s_1)^2 \, s_1 \, (1-s_2)^2 \, s_2} \, \log \frac{\mu}{m_b} +
\frac{\sum_{i=1}^{15} \hat{v}_i}{3 \, (1-s_1)^3 \, s_1 \, (1-s_2)^3 \, s_2} \right)
\label{virtualb} 
\end{eqnarray}
where the individual quantities $\hat{v}_1,\ldots,\hat{v}_{15}$ are relegated to
Appendix \ref{appendixvirt}.
\section{Bremsstrahlung corrections}
\label{sec:bremsstrahlung}
We now turn to the calculation of the bremsstrahlung QCD corrections, i.e. to
the contributions of order $\alpha_s$ with four particles in the
final state.
Before going into details, we mention that the kinematical range of
the variables $s_1$ and $s_2$ defined in
eq. (\ref{kinematicvariables}) is given  in this case
by\footnote{Strictly speaking, this range holds for $m_s=0$ and is
modified by powerlike terms of $m_s$, which we neglect in this paper.} 
$0 \le s_1 \le 1 \, ; \, 0 \le s_2 \le 1$. Nevertheless, we consider
in this paper only the range 
which is also accessible to the three-body decay $b \to s \gamma
\gamma$, i.e., $0 \le s_1 \le 1 \, ; \, 0 \le s_2 \le 1-s_1$ or, more
precisely, by its
restricted version specified in eq. (\ref{kinematical_cuts}).

\begin{figure}[h]
\begin{center}
\includegraphics[width=0.9\textwidth]{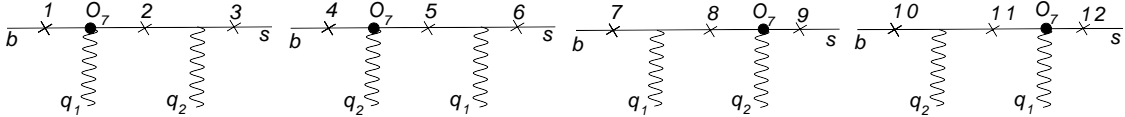}
\caption{The diagrams defining the
gluon-bremsstrahlung corrections to $b \to s \gamma \gamma$ are shown
at the amplitude level. The crosses in the graphs stand
for the possible emission places of the gluon.}
\label{fig:amplitudebrems}
\end{center}
\end{figure}

The diagrams defining the bremsstrahlung corrections at the
amplitude level are shown in Fig. \ref{fig:amplitudebrems}. 
The amplitude squared, needed to get the (double differential) decay width, can be written
as a sum of interferences of the different diagrams in
Fig. \ref{fig:amplitudebrems}. The four particle final state is
described by five independent kinematical variables (see section 
\ref{subsec:phasespace4}).

As already mentioned
in section \ref{sec:virtual}, the only source of the
singularities in the virtual corrections in our restricted range of
$s_1$ and $s_2$ is due to soft gluon-emission and/or
collinear emission of gluons from the $s$-quark.
When
analyzing the bremsstrahlung kinematics, one finds that there are situations
where one of the photons can become collinear with the
$s$-quark even within the mentioned restricted kinematical range of $s_1$
and $s_2$. While the singularities related to gluons cancel when
combining virtual- and bremsstrahlung corrections, those stemming from
collinear photon emission from the $s$-quark will remain and manifest
themselves as a term involving a single logarithm $\log(m_s)$ in the
final result.

In our previous paper \cite{Asatrian:2011ta} we realized that
for (formally) zero hadronic mass of the $(s,g)$-system collinear
photon emission is kinematically impossible. As a consequence, we
looked at the triple differential decay width 
$d\Gamma_{77}/(ds_1 ds_2 ds_3)$, where $s_3=(p_s+p_g)^2/m_b^2$ 
is the normalized hadronic
mass squared and found that the double differential decay width,
based on the triple differential decay width in which only the leading
power terms w.r.t. $s_3$ are retained, leads to a
nonsingular result when combined with the virtual corrections,
which we denoted by $d\Gamma_{77}^{\rm leading \, power}/(ds_1 ds_2)$ 
in ref. \cite{Asatrian:2011ta}.

In the present paper, working with a nonzero mass of the strange quark,
we go beyond leading power, keeping all terms which are independent of
$m_s$ and those which involve logarithms of $m_s$.

In the present paper we worked out in a first step the triple
differential spectrum $d\Gamma_{77}^{(1),brems}/(ds_1 ds_2 ds_3)$, 
for which we got a fully analytic result, which
is however rather lengthy. To get the double differential spectrum 
$d\Gamma_{77}^{(1),brems}/{ds_1 ds_2}$ we integrated over $s_3$, 
which runs in the
interval $[m_s^2/m_b^2,s_1 \cdot s_2]$. In some terms this integration was
done numerically. The final results (after combining with the virtual
corrections) are given in a form where certain parts have been fitted
to a set of $42$ ``basis function'', as the reader will see in the
following section.

As the details of the calculations are similar to those
in \cite{Asatrian:2011ta}, we refer to section 7 of that paper, 
where the used techniques are described in some detail.
In Appendix \ref{sec:phasespacemassive} we give, however, a useful
formula for the parametrization of the
$4$-particle phase-space for the case where one of the particles is massive. 

\section{Final result for the decay width at order $\alpha_s$}\label{sec:combination}
The complete order $\alpha_s$ correction to the
double differential decay width
$d\Gamma_{77}/(ds_1 \, ds_2)$ is obtained by
adding the renormalized virtual corrections from section
\ref{sec:virtual} and the bremsstrahlung corrections discussed in
section \ref{sec:bremsstrahlung}. 
We obtain (using $x_4=m_s^2/m_b^2$)
\begin{eqnarray}
&& \frac{d\Gamma_{77}^{(1)}}{ds_1 \, ds_2} = 
\frac{\alpha^2 \, \bar{m}_b^2(\mu) \, m_{b}^3 \, |C_{7,eff}(\mu)|^2 \, G_F^2 \,
  |V_{tb} V_{ts}^*|^2 \,  Q_d^2}{1024 \, \pi^5} \times \nonumber \\
&& \hspace{2cm} \frac{\alpha_s}{4\pi} \, C_F \,  \, 
\left[
\frac{-4 \, r_0 \, (1-s_1-s_2)}{(1-s_1)^2 \, s_1 \, (1-s_2)^2 \, s_2}
\, \log \frac{\mu}{m_b} +f + g \, \log(x_4) + h 
\right]  \, ,
\label{total}
\end{eqnarray}
where $r_0$ is given in eq. (\ref{funr0}). The first two terms in the
square bracket correspond to the leading power result, calculated in the
scheme where $m_s$ is different from zero, according to the present
paper. These two terms are exactly the same as in our previous paper 
\cite{Asatrian:2011ta} where the leading power terms where calculated
in the scheme with $m_s=0$. This coincidence, which has to hold of
course, provides a nontrivial check of our calculation. The remaining
two terms $g$ and $h$ encode all the non-leading power terms which are 
calculated for the first time in the present paper.
  
We now turn to the individual terms $f$, $g$ and $h$. As just
explained, $f$ is the same as in ref. \cite{Asatrian:2011ta} (see
eq. (5.2) there). For $g$ we obtain
\bea
g= && \frac{16 \, g_1 \, \log(s_1)}{s_1 (1+s_1)^3 (1-s_2)}  +
   \frac{16 \, g_2 \, \log(s_2)}{s_2 (1+s_2)^3 (1-s_1)}  +
   \frac{16 \, g_3 \, \log(1-s_1)}{s_2 (1+s_2)^3 }  + 
   \frac{16 \, g_4 \, \log(1-s_2)}{s_1 (1+s_1)^3 }  + \nonumber \\
&& \frac{16 \, g_5 \, (s_1+s_2) \, \log(s_1+s_2) +16 \, g_6 \,(1+s_1)
     \, (1+s_2)}{(1-s_1) s_1 (1+s_1)^3 (1-s_2) s_2 (1+s_2)^3 (s_1+s_2) }  
\eea
where the functions $g_1$,...,$g_6$ read
\bea
 g_1=&&-2 {s_1}^5-2 {s_2} {s_1}^4-{s_2}^2 {s_1}^3+2 {s_2}
   {s_1}^3+9 {s_1}^3+{s_2}^2 {s_1}^2+4 {s_2} {s_1}^2+ \nonumber \\
&& 17 {s_1}^2+8 {s_2} {s_1}+8 {s_1}+2 {s_2}^2+2
\eea
\bea
 g_2=&& g_1(s_1 \leftrightarrow s_2)
\eea
\bea
 g_3=&&-2 {s_1} {s_2}^4+6 {s_2}^4-4 {s_1} {s_2}^3+12 {s_2}^3-4
   {s_1} {s_2}^2+10 {s_2}^2+{s_1}
   {s_2}-{s_2}-{s_1}-1
\eea
\bea
 g_4=&& g_3(s_1 \leftrightarrow s_2)
\eea

\newpage

\bea
 g_5=&&-3 {s_2}^4 {s_1}^6-4 {s_2}^3 {s_1}^6-7 {s_2}^2 {s_1}^6-2
   {s_1}^6-4 {s_2}^5 {s_1}^5-3 {s_2}^4 {s_1}^5+ 8 {s_2}^3
   {s_1}^5+  \nonumber \\
&& 5 {s_2}^2 {s_1}^5-6 {s_1}^5-3 {s_2}^6 {s_1}^4-3
   {s_2}^5 {s_1}^4+36 {s_2}^4 {s_1}^4+58 {s_2}^3 {s_1}^4+31
   {s_2}^2 {s_1}^4- \nonumber \\
&& 15 {s_2} {s_1}^4-8 {s_1}^4-4 {s_2}^6
   {s_1}^3+8 {s_2}^5 {s_1}^3+58 {s_2}^4 {s_1}^3+64 {s_2}^3
   {s_1}^3+10 {s_2}^2 {s_1}^3- \nonumber \\ 
&& 32 {s_2} {s_1}^3-8 {s_1}^3-7
   {s_2}^6 {s_1}^2+5 {s_2}^5 {s_1}^2+31 {s_2}^4 {s_1}^2+10
   {s_2}^3 {s_1}^2-46 {s_2}^2 {s_1}^2- \nonumber \\
&& 35 {s_2} {s_1}^2-6
   {s_1}^2-15 {s_2}^4 {s_1}-32 {s_2}^3 {s_1}-35 {s_2}^2
   {s_1}-12 {s_2} {s_1}- 2 {s_1}- \nonumber \\ 
&& 2 {s_2}^6-6 {s_2}^5-8
   {s_2}^4-8 {s_2}^3-6 {s_2}^2-2 {s_2}
\eea
\bea
 g_6=&&4 {s_2}^4 {s_1}^6+{s_2}^3 {s_1}^6-4 {s_2}^2 {s_1}^6-5
   {s_2} {s_1}^6+4 {s_1}^6+8 {s_2}^5 {s_1}^5+9 {s_2}^4
   {s_1}^5 - \nonumber \\ 
&& 10 {s_2}^3 {s_1}^5-18 {s_2}^2 {s_1}^5+2 {s_2}
   {s_1}^5+9 {s_1}^5+4 {s_2}^6 {s_1}^4+9 {s_2}^5 {s_1}^4-12
   {s_2}^4 {s_1}^4- \nonumber \\ 
&& 29 {s_2}^3 {s_1}^4-14 {s_2}^2 {s_1}^4+22
   {s_2} {s_1}^4+8 {s_1}^4+{s_2}^6 {s_1}^3-10 {s_2}^5
   {s_1}^3-29 {s_2}^4 {s_1}^3- \nonumber \\ 
&& 28 {s_2}^3 {s_1}^3+3 {s_2}^2
   {s_1}^3+18 {s_2} {s_1}^3+5 {s_1}^3-4 {s_2}^6 {s_1}^2-18
   {s_2}^5 {s_1}^2-14 {s_2}^4 {s_1}^2+ \nonumber \\ 
&& 3 {s_2}^3 {s_1}^2+12
   {s_2}^2 {s_1}^2+3 {s_2} {s_1}^2+2 {s_1}^2-5 {s_2}^6
   {s_1}+2 {s_2}^5 {s_1}+22 {s_2}^4 {s_1}+ \nonumber \\ 
&& 18 {s_2}^3
   {s_1}+3 {s_2}^2 {s_1}+4 {s_2}^6+9 {s_2}^5+8 {s_2}^4+5
   {s_2}^3+2 {s_2}^2
\eea
The exact expression for the function $h$ in
eq. (\ref{total}) is very lengthy. We therefore write an
ansatz of the form
\bea
h&=&\frac{\sum_{i=1}^{42} c^h_i \, u_i}{(1-s_1)^3 s_1 (1-s_2)^3 s_2}
\, ,
\label{hfitted}
\eea
where the ``basis functions'' $u_i$ are given in
eq. (\ref{basisfunctions}) and where the coefficients $c^h_i$ (see Table
\ref{tab:coefffit})
are
obtained from a fit to the exact function $h$. 
For simpler use of our results and to make the present paper
self-contained, we also provide a fitted version for the
function $f$ according to
\bea
f&=&\frac{\sum_{i=1}^{42} c^f_i \, u_i}{(1-s_1)^3 s_1 (1-s_2)^3 s_2}
\, .
\label{ffitted}
\eea
The coefficients $c^f_i$ are also shown in Table \ref{tab:coefffit}.
We stress here that the fitted versions of $h$ and $f$ approximate the exact
functions very accurately in the whole phase-space, even when choosing the
parameter $c$ as small as $1/100$ (see eq. (\ref{kinematical_cuts})).

\newpage
 
The basis functions $u_i$ (which, like the exact
functions $h$ and $f$, are all
symmetric in $s_1$ and $s_2$) are chosen as 
\begin{eqnarray}
\nonumber
&&u_1=1,\quad u_2={s}_{1}+{s}_{2},\quad u_3={s}_{1}^2+{s}_{2}^2,\quad
u_4={s}_{1} {s}_{2},\quad u_5={s}_{1}^3+{s}_{2}^3,
\quad u_6= {s}_1^2 {s}_2+{s}_1 {s}_2^2, \\
\nonumber
&&u_7=\log ({s}_1)+\log ({s}_2),\quad 
u_8={s}_1\log ({s}_1)+{s}_2 \log ({s}_2),\quad
u_9={s}_2 \log ({s}_1)+{s}_1\log ({s}_2), \\
\nonumber
&&u_{10}={s}_1^2 \log ({s}_1)+{s}_2^2 \log ({s}_2),\quad
u_{11}={s}_1^2 \log ({s}_2)+{s}_2^2 \log ({s}_1), \\
\nonumber
&&u_{12}={s}_1 {s}_2 \log ({s}_1)+{s}_1 {s}_2 \log ({s}_2), \quad
u_{13}={s}_1^2 {s}_2 \log ({s}_1)+{s}_1 {s}_2^2 \log ({s}_2), \\
\nonumber
&&u_{14}={s}_1^2 {s}_2 \log ({s}_2)+{s}_1 {s}_2^2 \log ({s}_1),\quad
u_{15}={s}_1^3 \log ({s}_1)+{s}_2^3 \log ({s}_2),\\
\nonumber
&&u_{16}={s}_1^3 \log ({s}_2)+{s}_2^3 \log ({s}_1),\quad
u_{17}=\log ^2({s}_1)+\log ^2({s}_2), \\
\nonumber
&&u_{18}={s}_1 \log ^2({s}_1)+{s}_2 \log ^2({s}_2), \quad
u_{19}={s}_2 \log ^2({s}_1)+{s}_1 \log ^2({s}_2),\\
\nonumber 
&&u_{20}={s}_1^2 \log ^2({s}_1)+{s}_2^2 \log ^2({s}_2), \quad
u_{21}={s}_1^2 \log ^2({s}_2)+{s}_2^2 \log ^2({s}_1), \\
\nonumber 
&&u_{22}={s}_1 {s}_2 \log ^2({s}_1)+{s}_1 {s}_2 \log ^2({s}_2), \quad
u_{23}={s}_1^2 {s}_2 \log ^2({s}_1)+{s}_1 {s}_2^2 \log ^2({s}_2), \\
&&u_{24}={s}_1^2 {s}_2 \log ^2({s}_2)+{s}_1 {s}_2^2 \log ^2({s}_1), \quad
u_{25}={s}_1^3 \log ^2({s}_1)+{s}_2^3 \log ^2({s}_2),\\
\nonumber
&&u_{26}={s}_1^3 \log ^2({s}_2)+{s}_2^3 \log ^2({s}_1), \quad
u_{27}=\log ({s}_1) \log ({s}_2),\\
\nonumber
&&u_{28}=({s}_1+{s}_2) \log ({s}_1) \log ({s}_2),\quad
u_{29}=\left({s}_1^2+{s}_2^2\right) \log ({s}_1) \log ({s}_2),\\
\nonumber
&&u_{30}={s}_1 {s}_2 \log ({s}_1) \log ({s}_2),\quad
u_{31}=\left({s}_1^2 {s}_2+{s}_1 {s}_2^2\right) \log ({s}_1) \log
({s}_2),\\
\nonumber
&&u_{32}=\left({s}_1^3+{s}_2^3\right) \log ({s}_1) \log ({s}_2), \quad
u_{33}=\log (1-{s}_1)+\log (1-{s}_2),\\
\nonumber
&&u_{34}={s}_1 \log (1-{s}_1)+{s}_2 \log (1-{s}_2),\quad
u_{35}={s}_2 \log (1-{s}_1)+{s}_1 \log (1-{s}_2),\\
\nonumber
&&u_{36}={s}_1^2 \log (1-{s}_1)+{s}_2^2 \log (1-{s}_2), \quad
u_{37}={s}_1^2 \log (1-{s}_2)+{s}_2^2 \log (1-{s}_1),\\
\nonumber
&&u_{38}={s}_1 {s}_2 \log (1-{s}_1)+{s}_1 {s}_2 \log (1-{s}_2), \quad
u_{39}={s}_1^2 {s}_2 \log (1-{s}_1)+{s}_1 {s}_2^2 \log (1-{s}_2),\\
\nonumber
&& u_{40}={s}_1^2 {s}_2 \log (1-{s}_2)+{s}_1 {s}_2^2 \log (1-{s}_1), \quad
u_{41}={s}_1^3 \log (1-{s}_1)+{s}_2^3 \log (1-{s}_2),\\
\nonumber
&&u_{42}={s}_1^3 \log (1-{s}_2)+{s}_2^3 \log (1-{s}_1).
\label{basisfunctions}
\end{eqnarray}
\newpage
\begin{table}[h]
\centering \vspace{0.8cm}
\begin{tabular}{|c|c|c||c|c|c|}
\hline
$i$  & $c^f_i$ & $c^h_i$ & $i$ & $c^f_i$ & $c^h_i$
\\   \hline \hline
$1$ & $1587.9373 $ & $ 2808.0884 $ & $ 22 $ & $3839.3787$ & $8582.3121 $ \\ \hline 

$2$ & $-17820.810 $ & $ -27.836761 $ & $ 23$ & $2149.8019$ & $-3182.8383 $ \\ \hline 

$3$ & $5739.2134 $ & $ -127198.90 $ & $ 24$ & $-2969.4126$ & $-3814.5375 $ \\ \hline 

$4$ & $79681.671 $ & $ 150427.73 $ & $ 25$ & $1116.5578$ & $7876.6985 $ \\ \hline 

$5$ & $10672.929 $ & $ 123605.68 $ & $ 26$ & $-51.926335$ & $21.979815 $ \\ \hline 

$6$ & $-25630.099 $ & $ -61571.822 $ & $ 27$ & $-6.3461975$ & $0.42501969 $ \\ \hline 

$7$ & $206.57293 $ & $370.16329 $ & $ 28$ & $-198.78562$ & $243.20576 $ \\ \hline 

$8$ & $-6055.4090$ & $ -4884.2816 $ & $ 29$ & $-14.663373$ & $3294.2178 $ \\ \hline 

$9$ & $-1482.1360 $ & $261.69714 $ & $ 30$ & $-5234.3840$ & $-11486.898 $ \\ \hline 

$10$ & $-13734.475 $ & $ -59064.539 $ & $ 31$ & $-8078.6742$ & $-6953.1246 $ \\ \hline 

$11$ & $2458.1907 $ & $ 2819.9778 $ & $ 32$ & $463.51078$ & $1842.0350 $ \\ \hline 

$12$ & $2578.7004 $ & $ 19493.274 $ & $ 33$ & $-318.01486$ & $5524.1650 $ \\ \hline 

$13$ & $10698.372 $ & $ 29647.891 $ & $ 34$ & $1007.5887$ & $-13495.877 $ \\ \hline 

$14$ & $1305.9739 $ & $ 4481.7110 $ & $ 35$ & $17220.702$ & $9331.2971 $ \\ \hline 

$15$ & $-4990.6306 $ & $ -52868.520 $ & $ 36$ & $-1072.8013$ & $10698.386 $ \\ \hline 

$16$ & $-1135.5247$ & $ -3655.2789 $ & $ 37$ & $21912.257$ & $20102.580 $ \\ \hline 

$17$ & $17.550558 $ & $25.751857 $ & $ 38$ & $-29656.816$ & $-17993.661 $ \\ \hline 

$18$ & $-1255.7842 $ & $ -2016.3069 $ & $ 39$ & $12526.044$ & $ 8586.7318 $ \\ \hline 

$19$ & $-97.667743 $ & $ -87.275478 $ & $ 40$ & $-20491.027$ & $-18933.831 $ \\ \hline 

$20$ & $-755.27587 $ & $ -18097.634 $ & $ 41$ & $382.47503$ & $-2723.1301 $ \\ \hline 

$21$ & $135.25687 $ & $ 26.410005 $ & $ 42$ & $2606.0012$ & $2408.5233 $ \\ \hline 
\end{tabular}
  \caption{Coefficients $c^f_i$ and $c^h_i$, which occur in the fits of the functions $f$ and
    $h$, see eqs. (\ref{ffitted}) and (\ref{hfitted}).} 
\label{tab:coefffit}
\end{table}
The order $\alpha_s$ correction $d\Gamma_{77}^{(1)}/(ds_1 ds_2)$ in
Eq. (\ref{total}) to the double differential decay width
for $b \to X_s \gamma \gamma$ is the
main result of our paper.

\section{Numerical illustrations}\label{sec:numerics}
In the previous sections we calculated the virtual- and bremsstrahlung
QCD corrections associated with the operator ${\cal O}_7$.
While in the previous paper \cite{Asatrian:2011ta} only the leading
power terms in $s_3$ ($s_3$ is the normalized hadronic mass squared)
were taken into account in the underlying triple differential decay width
$d\Gamma_{77}/(ds_1 ds_2 ds_3)$, we performed a complete calculation in the
present paper. As there are configurations where one of the photons
can become collinear with the strange quark, we introduced a finite
mass $m_s$ which we consider to be of constituent type. While the
result based on leading power terms is finite in the limit $m_s \to
0$, the full result depends on $m_s$ through a single logarithm of
the form $\log(x_4)=\log(m_s^2/m_b^2)$. In the numerics we will vary
$m_s$ between $400$ MeV and $600$ MeV.

The NLL prediction reads
\be
\frac{d\Gamma_{77}}{ds_1 ds_2} = 
\frac{d\Gamma_{77}^{(0)}}{ds_1 ds_2} + \frac{d\Gamma_{77}^{(1)}}{ds_1 ds_2}
\label{total77}
\ee  
where the first- and second term of the r.h.s. are given in eqs. (\ref{treezero})
and (\ref{total}), respectively. 

\noindent
To illustrate our results, we first rewrite the $\overline{\mbox{MS}}$ mass
$\bar{m}_b(\mu)$ in eq. (\ref{total77}) in terms of the pole mass
$m_b$,
 using the one-loop relation
\[
\bar{m}_b(\mu) = m_b \, \left[ 1 - \frac{\alpha_s(\mu)}{4 \pi} \, \left( 8 \log
  \frac{\mu}{m_b} + \frac{16}{3} \right)
\right] \, .
\]
We then insert $C_{7,eff}(\mu)$ in the expanded form
(\ref{wilsonexpand}) and expand the resulting expression for
$d\Gamma_{77}/(ds_1 ds_2)$ w.r.t. $\alpha_s$, discarding terms of
order $\alpha_s^2$. This procedure defines the full NLL result
and also the version where only the leading power terms are retained
in $d\Gamma^{(1)}_{77}/(ds_1 ds_2)$. The
corresponding LL result is obtained by discarding the order
$\alpha_s^1$ term. The numerical values for
the input parameters and for this Wilson coefficient at various values
for the scale $\mu$, together with the
numerical values of $\alpha_s(\mu)$, are given in Table
\ref{tab:input} and Table \ref{tab:wilson}, respectively.

\begin{figure}[h]
\begin{center}
\includegraphics[width=\textwidth]{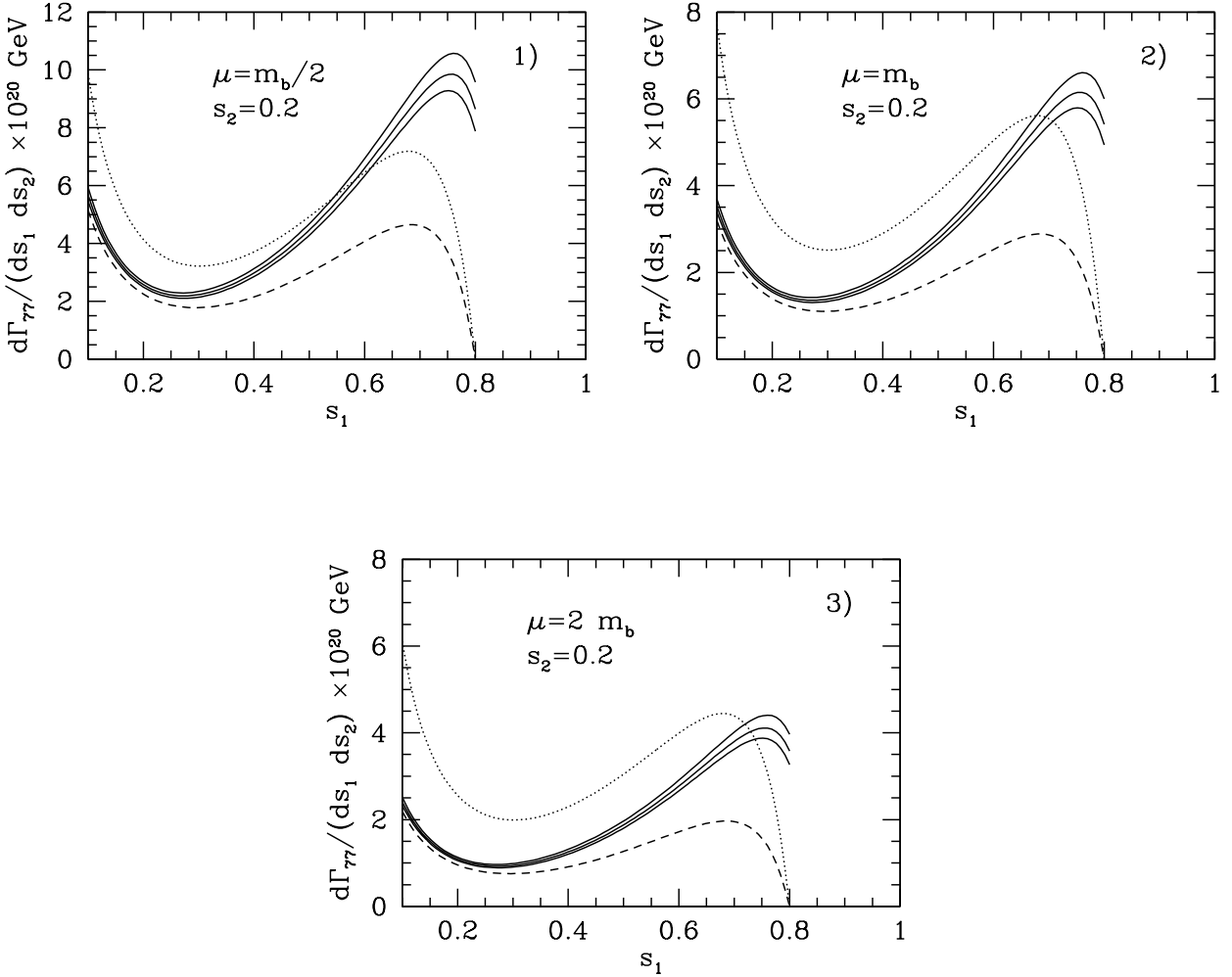}
\caption{Double differential decay width $d\Gamma_{77}/(ds_1 ds_2)$,
based on the operator ${\cal O}_7$ only, 
as a function of $s_1$ for $s_2$ fixed at $s_2=0.2$. 
The dotted, the dashed and the solid lines show 
the LL result, the NLL when only retaining leading power terms as in
ref. \cite{Asatrian:2011ta} and
the full NLL result of the present paper, respectively.  
Among the three solid lines, the highest, middle and
lowest curve correspond to $m_s=400$ MeV,  $m_s=500$ MeV and
$m_s=600$ MeV, respectively.
In the frames 1), 2) and 3) the renormalization scale is chosen to
be $\mu=m_b/2$, $\mu=m_b$ and $\mu=2 \, m_b$, respectively. See text
for details.}
\label{fig:results}
\end{center}
\end{figure}

In Fig. \ref{fig:results} the LL result, the NLL
result based on the leading power contribution and the full NLL result
are shown by the dotted, the dashed and the solid lines,
respectively. Among the three solid lines, the highest, middle and
lowest curve correspond to $m_s=400$ MeV,  $m_s=500$ MeV and
$m_s=600$ MeV, respectively.

From Fig. \ref{fig:results}, where $s_2$ is fixed at $s_2=0.2$, we see 
that for $s_1 \le 0.4$ the
NLL result is dominated by the leading power result obtained in our
previous paper \cite{Asatrian:2011ta}, while this is no longer true
for larger values of $s_1$. In these plots $s_1=0.8$ 
corresponds to the maximal value of the leading order
kinematics. In other words
the point $(s_1=0.8,s_2=0.2)$ lies on the ``diagonal line''
characterized by $1-s_1-s_2=0$ in Fig. \ref{fig:phasespace}.
That is why the dotted curves becomes zero at $s_1=0.8$. This also holds for the
virtual corrections which have the same kinematical range. 
The full kinematical range in the $(s_1,s_2)$-plane 
for the bremsstrahlung process is, however, larger than the window
considered in this paper. For this reason
the solid lines do not go to zero at $s_1=0.8$. However, the
leading power terms of the bremsstrahlung corrections have similar
features as the virtual corrections and go to zero for
$s_1=0.8$ (as seen from the dashed curves). A more detailed investigation shows
that the leading power contributions only give a good approximation of the
NLL result when one is sufficiently away from the line $1-s_1-s_2=0$
in the $(s_1,s_2)$-plane. 

The comparison of the full NLL corrections with the corresponding leading power
pieces is basically of ``historic'' interest; it is more
important to compare the LL curves (dotted) with the full NLL ones
(solid). Form Fig. \ref{fig:results} one
concludes that the NLL corrections to the ${\cal O}_7$  are crucial.
We stress that the QCD corrections involving 
the operators ${\cal O}_1$ and ${\cal O}_2$, which we did not consider
in our paper, also will be important.
Therefore, the issue concerning the
reduction of the $\mu$ dependence at NLL precision cannot be addressed
in a meaningful way at this level. 

To get the branching ratio for $\bar{B} \to X_s \gamma \gamma$ as a
function of the cut-off parameter $c$ defined in
eq. (\ref{kinematical_cuts}), we integrate the double differential
spectrum over the corresponding range in $s_1$ and $s_2$, divide by
the semileptonic decay width and multiply with the measured
semileptonic branching ratio. For the purpose of this paper it is
sufficient to take the lowest order formula for the semileptonic
decay width, reading
\bea
&& \Gamma_{sl} = \frac{m_b^5 \, G_F^2 \, |V_{cb}|^2}{192 \pi^3} \,
g(m_c/m_b) \, , 
\eea
with the phase space factor
\bea
&& g(z) = 1 - 8 \, z^2 + 8 \, z^6 - z^8 - 24 \, z^4 \log(z) \, .
\eea


%
%
Using the input parameters in Tables \ref{tab:input} and
\ref{tab:wilson}, we get the branching ratios shown in Table
\ref{tab:branching}
for the values $c=1/100$ (upper half) and $c=1/50$ (lower half)
at $\mu=m_b/2$, $\mu=m_b$ and $\mu=2 m_b$. In the columns ``${\cal
  O}_7$'' 
only the operator ${\cal O}_7$ is taken into account, while the number
in the columns ``all'' also takes into account the lowest order
contributions involving the operators ${\cal O}_1$ and ${\cal O}_2$
(according to eq. (\ref{treeallzero})).

\begin{table}[h]
\centering \vspace{0.8cm}
\begin{tabular}{|c|c|c|c|c|c|c|}
\hline
   & ${\cal O}_7$ & all & ${\cal O}_7$ & all & ${\cal O}_7$ & all
\\   \hline 
   & $\mu=m_b/2$ & $\mu=m_b/2$ & $\mu=m_b$ & $\mu=m_b$ & $\mu=2m_b$ &$\mu=2 m_b$
   \\ \hline 
$\mbox{LL}$    & 3.96& 3.96 & 3.10& 3.11 &2.45 & 2.53 \\ \hline
$\mbox{NLL}_1$ & 3.81& 3.81 & 2.37& 2.39 &1.60 & 1.68 \\ \hline
$\mbox{NLL}_2$ & 3.35& 3.34 & 2.08& 2.10 &1.41 & 1.49 \\ \hline
$\mbox{NLL}_3$ & 2.97& 2.97 & 1.85& 1.87 &1.25 & 1.33 \\ \hline \hline

$\mbox{LL}$    & 2.40& 2.40 & 1.87& 1.89 &1.48 & 1.55 \\ \hline
$\mbox{NLL}_1$ & 2.39& 2.39 & 1.49& 1.51 &1.01 & 1.08 \\ \hline
$\mbox{NLL}_2$ & 2.17& 2.17 & 1.35& 1.37 &0.91 & 0.99 \\ \hline
$\mbox{NLL}_3$ & 1.99& 1.99 & 1.24& 1.26 &0.84 & 0.91 \\ \hline 
\end{tabular}
  \caption{Branching ratios for $\bar{B} \to X_s \gamma \gamma$ in
    units of $10^{-7}$. The upper half of the table is for $c=1/100$
    and lower half for $c=1/50$. LL is the leading logarithmic
    result. $\mbox{NLL}_1$,  $\mbox{NLL}_2$ and $\mbox{NLL}_3$
    are the results where the NLL corrections to the
    ${\cal O}_7$ contributions are included, using $m_s=400$ MeV, 
    $m_s=500$ MeV and $m_s=600$ MeV, respectively. See text for 
more information.}
  \label{tab:branching}
\end{table}


\section{Summary}\label{sec:summary}
In the present work we calculated the $O(\alpha_s)$
corrections to the decay process $\bar{B} \to X_s \gamma \gamma$
originating from diagrams involving the electromagnetic
dipole operator ${\cal O}_7$. This calculation involves contributions
with three particles in the final state and a gluon in the loop (virtual
corrections) and tree-level contributions with
four particles in the final state (gluon bremsstrahlung corrections).

We introduced a nonzero mass $m_s$ for the strange quark to regulate
configurations where the gluon or one of the photons become collinear
with the strange quark and retained terms which are logarithmic in
$m_s$, while discarding terms which go to zero in the limit $m_s \to 0$.
When combining virtual- and bremsstrahlung corrections, the infrared
and collinear singularities induced by soft and/or collinear gluons
drop out. By our cuts the photons do not become soft, but one of them can
become collinear with the strange quark. This implies that in the
final result a single logarithms of $m_s$ survives. We interpret $m_s$
appearing in the result as a constituent mass and vary it between
$400$ MeV and $600$ MeV in the numerics. 

We find that the NLL corrections to the double differential spectrum
$d\Gamma_{77}/(ds_1 ds_2)$ are large in general. Depending on the
point in the $(s_1,s_2)$-plane, they can modify the LL predictions
by up to $50\%$ in both directions, which means that not only the
normalization, but also the shapes of the distributions are modified,
as can be seen e.g. in Figure \ref{fig:results}.
 
We also compared our new results with those obtained in an earlier
paper \cite{Asatrian:2011ta}, where only the leading power terms w.r.t. $s_3$
in the underlying triple differential spectrum $d\Gamma_{77}/(ds_1 ds_2
ds_3)$ were retained.

\section*{\normalsize Acknowledgments}
\vspace*{-2mm}
C.G. was supported by the Swiss National Science Foundation.
He also thanks for the kind hospitality extended to him by the DESY
Theory group during his Sabbatical, where a part of this work was done.

\noindent
H.M.A. was also supported by the Swiss National Science Foundation, the
Volkswagen Stiftung Program No. 86426 and the State Committee of
Science of  Armenia Program No. 13-1C153.

\noindent
We thank Ahmed Ali for useful discussions and carefully reading the
manuscript.

\newpage

\appendix
\section{Explicit results for the functions $\hat{v}_i$ defining the
virtual corrections}
\label{appendixvirt}
The functions $\hat{v}_i$ appearing in eq. (\ref{virtualb}) read
\begin{eqnarray}
\nonumber \hat{v}_1&=&16 (1-s_1-s_2) \left[\left(96-2 \pi ^2\right)
   s_1^4 s_2^4+\left(11 \pi ^2-291\right)
   s_1^4 s_2^3+\left(300-19 \pi ^2\right)
   s_1^4 s_2^2 + \right. \\  \nonumber
   &&\left. \left(12 \pi ^2-117\right)
   s_1^4 s_2+\left(12-3 \pi ^2\right)
   s_1^4+\left(11 \pi ^2-291\right) s_1^3
   s_2^4+\left(894-36 \pi ^2\right) s_1^3
   s_2^3 + \right.  \\ \nonumber
   &&\left. \left(48 \pi ^2-936\right) s_1^3
   s_2^2+\left(348-21 \pi ^2\right) s_1^3
   s_2+\left(2 \pi ^2-15\right)
   s_1^3+\left(300-19 \pi ^2\right) s_1^2
   s_2^4  + \right. \\ \nonumber
   &&\left. \left(48 \pi ^2-936\right) s_1^2
   s_2^3+\left(1044-60 \pi ^2\right) s_1^2
   s_2^2+\left(26 \pi ^2-426\right) s_1^2
   s_2+\left(18-\pi ^2\right) s_1^2 - \right. \\ \nonumber
   &&\left.\pi ^2\left(s_1^5 s_2^3-3 s_1^5
   s_2^2+3
   s_1^5 s_2-s_1^5+s_1^3
   s_2^5-3 s_1^2 s_2^5+3 s_1
   s_2^5-s_2^5\right) + \right. \\ \nonumber
   &&\left. \left(12 \pi ^2-117\right)
   s_1 s_2^4+\left(348-21 \pi ^2\right) s_1
   s_2^3+\left(26 \pi ^2-426\right) s_1
   s_2^2+\left(210-14 \pi ^2\right) s_1
   s_2 + \right. \\ \nonumber
   &&\left.\left(\pi ^2-15\right) s_1+\left(12-3
   \pi^2\right) s_2^4+\left(2 \pi ^2-15\right)
   s_2^3+\left(18-\pi ^2\right) s_2^2+\left(\pi
   ^2-15\right) s_2\right]
 \end{eqnarray}
 \begin{eqnarray}
\nonumber \hat{v}_2&=&   -96 s_1 s_2 \left(1-s_1\right)^3 \left(1-s_2\right)^2
   \left(1-s_1-s_2\right) \left( 2 - 3 s_2 \right) \,
 \log \left(s_1\right)
 \end{eqnarray}
 \begin{eqnarray}
\nonumber 
\hat{v}_3&=& 48 s_1 s_2 \left(1-s_1\right)^2 \left(1-s_2\right)^2
   \left(1-s_1-s_2\right) \left(s_1 -s_1^2 +2 s_2 - s_1 s_2 \right)
   \log ^2\left(s_1\right)
 \end{eqnarray}
\begin{eqnarray}
\nonumber \hat{v}_4&=&
-96 \left(1-s_1\right){}^2
   \left(1-s_2\right){}^2 s_2 \left(s_1^4+2
   s_2 s_1^3-2 s_1^3+s_2^2 s_1^2-4 s_2
   s_1^2+s_1^2-2 s_2^2 s_1+
\right.  \\&& \left.  \nonumber   
   3 s_2 s_1-2
   s_1+s_2^2+1\right) \log \left(s_1\right)
   \log \left(s_1+s_2\right)\end{eqnarray}
\begin{eqnarray}
\nonumber \hat{v}_5&=&
48 \left(1-s_1\right) \left(s_2-1\right){}^2
   s_2 \left(1-s_1-s_2\right) \left(6 s_2
   s_1^3-6 s_1^3-11 s_2 s_1^2+15 s_1^2+
   \right.  \\&& \left.  \nonumber
   3 s_2 s_1-9 s_1+2\right) \log \left(1-s_1\right)
   \end{eqnarray}
\begin{eqnarray}
\nonumber \hat{v}_6&=&
96 \left(1-s_1\right) \left(s_2-1\right){}^2
   \left(s_2 s_1^5-s_1^5+2 s_2^2 s_1^4-5 s_2
   s_1^4+3 s_1^4+s_2^3 s_1^3-5 s_2^2 s_1^3+
   \right.  \\&& \left.  \nonumber
   8 s_2 s_1^3-2 s_1^3-s_2^3 s_1^2+4 s_2^2
   s_1^2-4 s_2 s_1^2+s_1^2-4 s_2^2 s_1+3 s_2
   s_1-s_1-s_2^2+s_2\right) \times
   \\&&   \nonumber
   \log\left(1-s_1\right) \log
   \left(s_1+s_2\right)
   \end{eqnarray}
\begin{eqnarray}
\nonumber \hat{v}_{7}&=&48 \left(1-s_1\right) \left(1-s_2\right)
   \left(s_2^2 s_1^5-s_2 s_1^5-9 s_2^3
   s_1^4+16 s_2^2 s_1^4-8 s_2 s_1^4+s_1^4-9
   s_2^4 s_1^3+
 \right.  \\&& \left.  \nonumber   
   46 s_2^3 s_1^3-67 s_2^2
   s_1^3+35 s_2 s_1^3-s_1^3+s_2^5 s_1^2+16
   s_2^4 s_1^2-67 s_2^3 s_1^2+84 s_2^2
   s_1^2-
   \right.  \\&& \left.  \nonumber 
   43 s_2 s_1^2+s_1^2-s_2^5 s_1-8 s_2^4
   s_1+35 s_2^3 s_1-43 s_2^2 s_1+22 s_2
   s_1-s_1+s_2^4-s_2^3+
   \right.  \\&& \left.  \nonumber 
   s_2^2-s_2\right) \log
   ^2\left(s_1+s_2\right)
\end{eqnarray}
\begin{eqnarray}
\nonumber \hat{v}_{8}&=&96 s_1 \left(1-s_2\right){}^2
   \left(1-s_1-s_2\right) \left(s_2
   s_1^4-s_1^4+s_2^2 s_1^3-4 s_2 s_1^3+3
   s_1^3-5 s_2^2 s_1^2+
   \right.  \\&& \left.  \nonumber
   8 s_2 s_1^2-2 s_1^2+7
   s_2^2 s_1-11 s_2 s_1+s_1-2 s_2^2+5
   s_2-1\right) \text{Li}_2\left(s_1\right)
\end{eqnarray}
\begin{eqnarray}
\nonumber \hat{v}_{9}&=&96 \left(1-s_1\right) \left(1-s_2\right)
   \left(1-s_1-s_2\right) \left(s_2^2 s_1^4-2
   s_2 s_1^4+s_1^4+8 s_2^3 s_1^3-17 s_2^2
   s_1^3+
   \right.  \\&& \left.  \nonumber
   12 s_2 s_1^3-3 s_1^3+s_2^4 s_1^2-17
   s_2^3 s_1^2+32 s_2^2 s_1^2-20 s_2 s_1^2-2
   s_2^4 s_1+12 s_2^3 s_1-
   \right.  \\&& \left.  \nonumber
   20 s_2^2 s_1+20 s_2
   s_1-2 s_1+s_2^4-3 s_2^3-2 s_2\right)
   \text{Li}_2\left(1-s_1-s_2\right)
\end{eqnarray}
\begin{eqnarray}
&& \hat{v}_{10}=\hat{v}_2(s_1\leftrightarrow s_2)\quad \quad
 \hat{v}_{11}=\hat{v}_3(s_1\leftrightarrow s_2)\quad \quad 
 \hat{v}_{12}=\hat{v}_{4}(s_1\leftrightarrow s_2)\quad \quad
 \nonumber \\
 && \hat{v}_{13}=\hat{v}_{5}(s_1\leftrightarrow s_2)\quad \quad
 \hat{v}_{14}=\hat{v}_{6}(s_1\leftrightarrow s_2)
 \quad \quad
 \hat{v}_{15}=\hat{v}_{8}(s_1\leftrightarrow s_2)
\end{eqnarray}
\section{Relevant phase-space formulas}
\label{sec:phasespacemassive}
The fully differential decay width $d\Gamma$ for a generic process 
$p \to p_1 + p_2 + ... + p_n$ can be written as
\bea
d\Gamma = \frac{1}{2m} \, \overline{|M|^2} \, D\Phi(1 \to n) \, ,
\eea
where  $\overline{|M|^2}$ is the squared matrix element, summed and
  averaged
over spins and colors of the particles in the final and initial
state, respectively, and $m$ is the mass of the decaying particle.

In ref. \cite{Asatrian:2012tp} useful parametrizations for the
phase-space factors $D\Phi(1 \to n)$ have been given for $n=3,4$,
for the case when all final-state particles are massive. Among the
final-state particles only the strange quark is massive in our
application, which means that the general formulas simplify.
In the following subsections we see that the 3-particle
phase-space can be parametrized in terms of two parameters
$\lambda_1$ and $\lambda_2$, which run independently in the range
$[0,1]$, while five such parameters ($\lambda_1,...,\lambda_5$) are 
involved in the 4-particle phase-space. Of course, all scalar products involved
in  $\overline{|M|^2}$ can be expressed in terms of these parameters.
\subsection{Phase-space parametrization for the 3-particle final
  state}
\label{subsec:phasespace3}
In our application we identify $p_1$ with the strange quark and $p_2,
p_3$ with the two photons and define $x_1=m_s^2/m_b^2$. Starting from
eq. (2.10) of ref. \cite{Asatrian:2012tp}, one gets
\begin{eqnarray}
D\Phi(1\to 3)= && \frac{{m}_b^{2 d-6}2^{1-2 d} \pi^{1-d}}{\Gamma(d-2)}
 [(1-\lambda_1)
   \lambda_1]^{\frac{d-4}{2}} [(1-\lambda_2)
   \lambda_2]^{d-3} \times \nonumber \\
&& (1-{x}_1)^{2d-5} [\lambda_2
   (1-{x}_1)+{x}_1]^{\frac{2-d}{2}} \, d\lambda_1 d\lambda_2 \, .  
\end{eqnarray}
The scalar products of the momenta $p_i$, encoded in the quantities
$s_{ij}=(p_i+p_j)^2/m_b^2$, can be written in terms of the
parameters $\lambda_1$ and $\lambda_2$ as
\begin{eqnarray}
s_{13}&=&\lambda_2 (1-{x}_1)+{x}_1 \nonumber\\
s_{12}&=&\frac{\lambda_1 (\lambda_2-1) \lambda_2
   (1-{x}_1)^2-{x}_1}{\lambda_2
   ({x}_1-1)-{x}_1}\nonumber \, .
\end{eqnarray}
From the observation that $s_1=s_{13}$ and $s_2=s_{12}$ one easily
gets the expression for the double differential spectrum
$d\Gamma/(ds_1 ds_2)$.

\subsection{Phase-space parametrization for the 4-particle final
  state}
\label{subsec:phasespace4}
In our application we identify $p_1,p_2$ with the two photons, $p_3$
with the gluon and $p_4$ with the strange quark
and define $x_4=m_s^2/m_b^2$. Starting then from eq. (3.10) of ref.
\cite{Asatrian:2012tp}, putting there $x_1=x_2=x_3=0$ and performing
the substitutions
\begin{eqnarray}
\nonumber
&&z_1=2 \lambda_3 - 1, \quad {z}_{32}=2 \lambda_5 - 1,\quad 
{s}_{234}=\lambda_1 (1-{x}_4)+{x}_4 \, \\
&&E_2=\frac{\lambda_1 (1-\lambda_2) (1-{x}_4)}{2 \sqrt{
   \lambda_1+{x}_4-\lambda_1 {x}_4}} \, , \\
\nonumber
&&{z}_{31}= \frac{\lambda_1 (1-{x}_4) (\lambda_2
   (1-\lambda_4)-\lambda_4)+(1-2 \lambda_4)
   {x}_4}{\lambda_1 (1-{x}_4) (\lambda_2
   (1-\lambda_4)+\lambda_4)+{x}_4}   \, ,
\end{eqnarray}
we get the following expression for the phase-space factor:
\begin{eqnarray}
\nonumber D\Phi(1\to 4)&=&(4\pi)^{-\frac{3d}{2}}m_b^{3d-8}
\frac{2^{2d-7}\Gamma(\frac{d-2}{2})}{(d-3)\Gamma(d-3)^2}
 (1-{x}_4)^{3 d-7}  [(1-\lambda_1) (1-\lambda_2)
   \lambda_2]^{d-3}\\
   &\times &\lambda_1^{2 d-5}[(\lambda_1
   (1-{x}_4)+{x}_4) (\lambda_1 \lambda_2
   (1-{x}_4)+{x}_4)]^{1-\frac{d}{2}}\\
  &\times &\nonumber [(1-\lambda_3) \lambda_3 (1-\lambda_4)
   \lambda_4]^{\frac{d}{2}-2} 
    [(1-\lambda_5) \lambda_5]^{\frac{d-5}{2}} \, d\lambda_1
d\lambda_2 d\lambda_3 d\lambda_4 d\lambda_5 \, .
   \end{eqnarray}
As mentioned above, all $\lambda_i$ run independently in the range
$[0,1]$.
All scalar products of the momenta $p_i$, encoded in the quantities
$s_{ij}=(p_i+p_j)^2/m_b^2$ and $s_{ijk}=(p_i+p_j+p_k)^2/m_b^2$, 
can be written in terms of the
parameters $\lambda_1,...,\lambda_5$ as
\begin{eqnarray}
\nonumber s_{234}&=&\lambda_1(1-{x}_4)+{x}_4 \, , \\
\nonumber s_{34}&=&\lambda_1\lambda_2(1-{x}_4)+{x}_4 \, , \\
s_{23}&=&\frac{\lambda_1^2 (1-\lambda_2) \lambda_2 \lambda_4
   (1-{x}_4)^2}
   {\lambda_1 \lambda_2(1-{x}_4)+{x}_4} \, , \\
\nonumber s_{134}&=&\frac{\lambda_1 (1-{x}_4) [\lambda_2(1- (1-\lambda_1)
   \lambda_3 (1-{x}_4))+\lambda_3 (1-\lambda_1)(1-{x}_4)]+{x}_4}{\lambda_1
   (1-{x}_4)+{x}_4} \, , \\
\nonumber s_{13}&=&(s_{13}^{+}-s_{13}^{-})\lambda_5+s_{13}^{-} \, ,
\end{eqnarray}
where
\begin{eqnarray}
\nonumber s_{13}^{\pm}&=&\frac{(1-\lambda_1) \lambda_1 \lambda_2 (1-{x}_4)^2} 
{(\lambda_1+{x}_4-\lambda_1 {x}_4) 
(\lambda_1 \lambda_2+{x}_4-\lambda_1
   \lambda_2 {x}_4)}
\left\{ {x}_4 [(1-\lambda_3)
   (1-\lambda_4)+\lambda_3 \lambda_4]\right.\\ &+& \left.
   (1-{x}_4) \lambda_1 [\lambda_2 (1-\lambda_3) (1-\lambda_4)
  + \lambda_3 \lambda_4] \right.\\ 
 &\mp&\left.  2\sqrt{(1-\lambda_3) \lambda_3(1-\lambda_4) 
\lambda_4 (\lambda_1+{x}_4-\lambda_1{x}_4) 
(\lambda_1
   \lambda_2+{x}_4-\lambda_1 \lambda_2 {x}_4)}
   \nonumber \right\} \, .
\end{eqnarray}
From the observation that $s_1=s_{234}$, $s_2=s_{134}$ and
$s_3=s_{34}$ one 
easily
gets the expression for the triple differential spectrum
$d\Gamma/(ds_1 ds_2 ds_3)$.

\section{Renormalization constants}\label{append:renormalizationconstants}
In this appendix, we collect the  explicit expressions of the renormalization constants needed for the
ultraviolet renormalization in our calculation (see section \ref{sec:virtual}).

\noindent The operator ${\cal O}_{7}$, as well as the $b$-quark mass
contained in this operator are renormalized in the $\MS$ scheme \cite{Misiak:1994zw}:
\be
 Z_{77}^{\MS} = 1 + \frac{4\,C_F}{\epsilon}\frac{\alpha_s(\mu)}{4\pi}
 + O(\alpha_s^2) \quad ; \quad
 Z_{m_b}^{\MS} = 1 - \frac{3\,C_F}{\epsilon}\frac{\alpha_s(\mu)}{4\pi}
 + O(\alpha_s^2)\, .
\ee

\noindent All the remaining fields and parameters are 
renormalized in the on-shell scheme. The on-shell renormalization constant for
the $b$-quark mass is given by
\be
 Z_{m_b}^{\rm OS} =
 1-C_F\,\Gamma(\epsilon)\,e^{\gamma\epsilon}\,
 \frac{3-2\epsilon}{1-2\epsilon}
 \left(\frac{\mu}{m_b}\right)^{2\epsilon}\frac{\alpha_s(\mu)}{4\pi} +
 O(\alpha_s^2)\, .
\ee
while the renormalization constants for the  $s$- and 
$b$-quark fields are ($q=b$ or $q=s$)
\bea
  Z_{2q}^{\rm OS} &=& 1 - C_F\,\Gamma(\epsilon)\,e^{\gamma\epsilon}\,\frac{3-2\epsilon}{1-2\epsilon}
  \left(\frac{\mu}{m_q}\right)^{2\epsilon}\frac{\alpha_s(\mu)}{4\pi} +
  O(\alpha_s^2)\, .
\eea
The various quantities $\delta Z$ appearing in section
\ref{sec:virtual} are defined to be $\delta Z = Z-1$.

\end{document}